\documentclass[aps,twocolumn]{revtex4-1}
\usepackage{amsmath,amssymb}
\usepackage{slashed}
\usepackage{graphicx}
\usepackage{bm}
\usepackage[T1]{fontenc}
\usepackage[utf8]{inputenc}
\usepackage{gauss} 
\usepackage[top=1.0in,bottom=1.0in,left=1.0in,right=1.0in]{geometry}
\usepackage[colorlinks,linkcolor=blue,citecolor=blue]{hyperref}
\usepackage{comment}
\usepackage{ytableau}
\newcommand{\be}{\begin{equation}}
\newcommand{\ee}{\end{equation}}
\newcommand{\bea}{\begin{eqnarray}}
\newcommand{\eea}{\end{eqnarray}}
\newcommand{\ba}{\begin{eqnarray}}
\newcommand{\ea}{\end{eqnarray}}

\begin{document}

\title{Quantum potential with no  perturbative series, \\
and nonperturbative vacuum  dominated by complex  classical paths
}

\author{Edward Shuryak}
\email{edward.shuryak@stonybrook.edu}
\affiliation{Center for Nuclear Theory, Department of Physics and Astronomy, Stony Brook University, Stony Brook, New York 11794--3800, USA}


\begin{abstract} 
Spectra of standard 1d potentials (double-well,
sin-Gordon etc) are given by trans-series in coupling, including (badly
divergent) perturbative series (PS), and 
nonperturbative terms $\sim exp(-const/g^2)$ and $log(1/g)exp(-const/g^2)$. All of them are badly defined (e.g. PS are badly
divergent) but in sum supposed to be good.
In this paper we discuss an example of a potential with specially defined couplings making PS completely absent. We calculate its nonperturbative vacuum energy and show that they are reproduced by the action of 
 certain complex solutions to holomorphic Newton equation. 
\end{abstract}

\maketitle
\section{Introduction}

Perhaps the most spectacular early example of how Quantum Mechanics is different from the classical one was semiclassical description of tunneling \cite{Gamow:1928zz}. It spectacularly described  {\em $\alpha$ decays}
and spontaneous fissions of heavy nuclei, covering the range of probabilities
spanning about 20 orders of magnitude.

Instanton, a space-time soliton in gauge theories \cite{Belavin:1975fg},
plays a very important role in gauge dynamics. As was shown in my early
``instanton liquid model" (ILM)
\cite{Shuryak:1981ff} and confirmed
in many subsequent work, instantons 
create spontaneous breaking of the chiral symmetry in QCD, being thus responsible for effective quark mass
(and thus, the mass of nucleons and ourselves). Lately, numerical
simulations on the lattice and phenomenological studies also
point to important role of correlated instanton-antiinstanton pairs, or molecules. For recent review see e.g.
\cite{Shuryak:2026pqt}.

Very extensive developments of
semiclassical theory, based on Feynman path integrals
 in imaginary time, was slowly developed in the meantime.  
The $instantons$ were discussed in many more settings. The most studied is  the {\em double well potential} (DWP) with a potential
\be V_{DWP} \sim (x^2-f^2)^2
\ee
with two degenerate minima, at $x=\pm f$.  Tunneling between them lifts degeneracy of the spectrum
and establishes parity ($x\leftrightarrow -x$) as the true
quantum number. 

\begin{figure}[b]
    \centering   \includegraphics[width=0.85\linewidth]{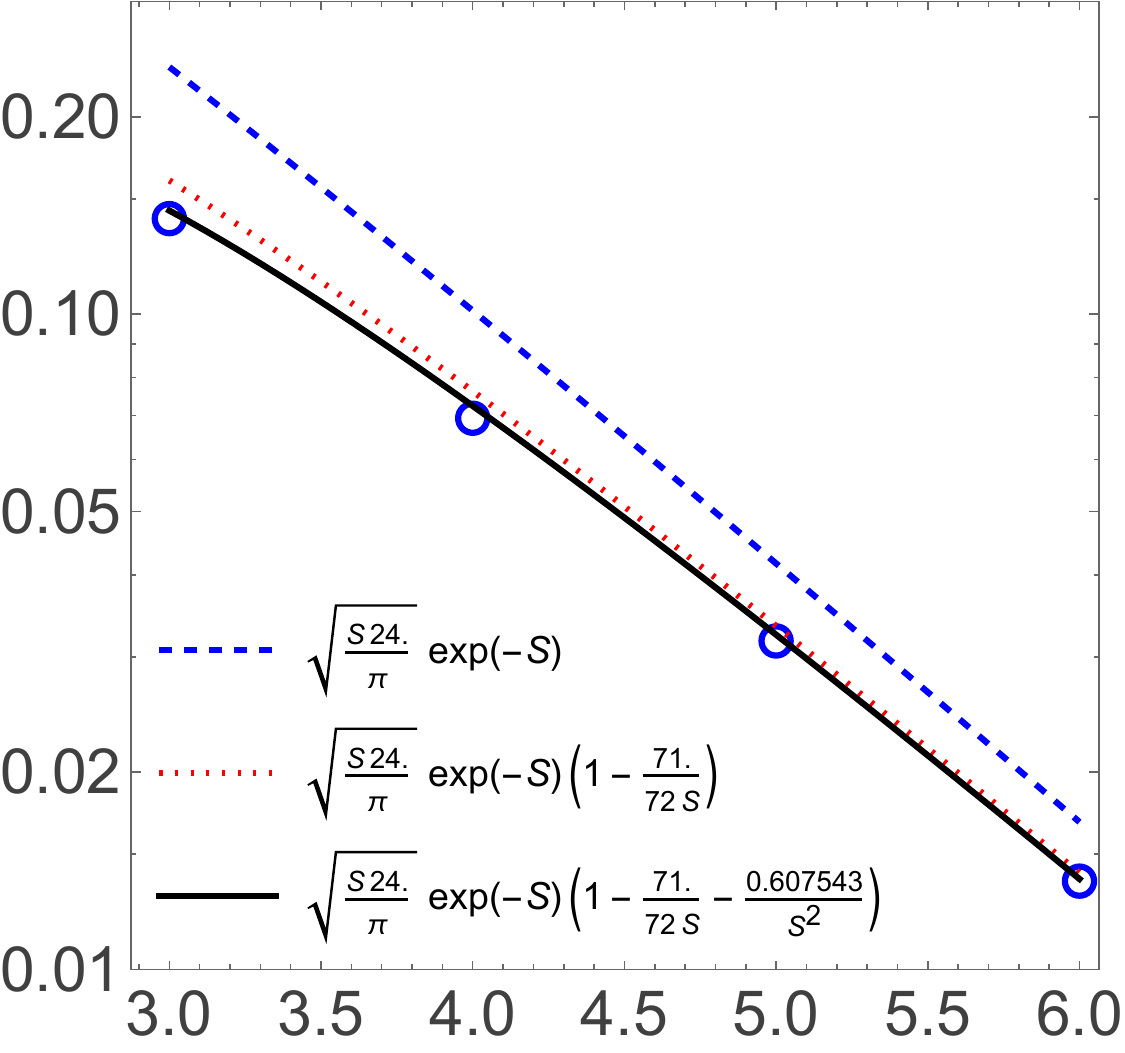}
    \caption{The splitting of the negative-parity state from the vacuum,
    $gap=E_- - E_+$, shown as data points. The three curves correspond to
    one-, two-, and three-loop semiclassical approximations.}
    \label{fig_DWP_gap_semi}
\end{figure}

Further developments of semiclassical theory was formulating it in QFT terms.
Using Feynman diagrams allows in principle calculate corrections 
to the instanton amplitude to
any loop order.
In particular, corrections up to three loops have been calculated 
 \cite{Escobar-Ruiz:2015nsa}. 
 In Fig.~\ref{fig_DWP_gap_semi} we compare the one-, two-, and three-loop semiclassical expressions to
actual $gaps$ (splitting between lowest negative-to-positive parity states) $E_- - E_+$ computed numerically and given as  a function of the instanton action $S$. This plot demonstrate that,
with fluctuations to few loop included, semiclassical theory becomes rather accurate, even at not-so-large values of the action $S\sim 3\hbar$.
 
A wider perspective on semiclassics
relates it to old questions of mathematics on representing solutions as series in parameters.
Dependence of physical observables 
(such as the vacuum energy we will
mostly discuss below) on coefficients of the action
should in general be 
expressed as the so called
{\em trans-series}. Those combine
(badly divergent) perturbative series ($\sum a_n g^n$) with nonperturbative terms ($\sim exp(-const/g^2)$ times another
series  $ \sum b_n g^n$), for 
a review see e.g. \cite{Dunne:2014bca}. Nontrivial
relations between $a_n$ and $b_n$
coefficients is known as $resurgence$ relations. While those are fulfilled in known examples (e.g. by
two and three-loop corrections just mentioned) their generality, or even existence in the QFT context, remains a mystery.

Since early 1980's, a number of boson-fermion examples were
given in the context of
   {\em supersymmetric quantum mechanics} 
\cite{Witten:1982df}. In those  
 supersymmetry holds perturbatively but gets broken nonperturbatively, by instantons. One example
 \cite{Turbiner:1991np}, with potential
\be V_{T0}=a^2 x^6 - 2a x^4 + (1+3a)x^2 -1 \ee
 was studied extensively, showing that vacuum energy is indeed 
 exponentially small
$\sim ~ a^{1/2} \exp {- const /a} $
at small $a$. 

In current paper we focus
on other special 1d QM Hamiltonian
\be  \label{eqn_pot}
H=(p^2+x^2-1)+(a^2 x^4 +2a x^3-2 a x)
\ee
suggested to us by A.Turbiner (private communication). 

First of all, we will prove in the next section
a total vanishing of
perturbative series
(all 
$a_n=0$), done without any use of fermions or supersymmetry.
Yet the noperturbative effects still apparently exist, and our further study would be to look
for semiclassical paths responsible for the vacuum energy.

In doing so, we will follow 
 \cite{Behtash:2015zha}, 
 who introduced a notion of ``complex bions", solutions of
he {\em holomorphic Newton’s equation} (HNE). Let us remind that this work was set for 
  {\em tilted double well} (TDW) with  potentials 
\ba  V_\pm(x)&=&{1 \over 2}(W'(x))^2  \pm {p\over 2}W''(x),\nonumber \\
W(x) &=& x^3/3-x
\ea
and parameter $p$ close to 1.

 Flipping the potential ($V\rightarrow -V$) in Euclidean time formulation,  one observes that real paths starting from the
 global maximum cannot have a finite action. There exist however complex paths,  solutions to the {\em holomorphic Newton’s equation} (HNE) with finite action. A solution of such kind,
 or {\em complex bion}, was found in
 \cite{Behtash:2015zha} by analytic continuation. Among other
  important observation made by \cite{Behtash:2015zha} let us mention that HNE separated for real and imaginary
parts do $not$ correspond to motion in
any 2d potential.

(Another extreme path at small $a$ can be a real
"bounce" path, starting from the lower maximum and reflected from real turning point. 
 Yet it is of no interest for us now, as it
is related to the false vacuum problem and is irrelevant
to current work.)


\subsection{Outline of the paper}
 
In this work we focus on a specific  example of 1d QM problem with Hamiltonian
(\ref{eqn_pot}).
Note a bit nonstandard selection of units in the first Harmonic bracket,
without usual 1/2 and with some nonzero vacuum energy subtracted. 

Three other terms, $\sim x,x^3,x^4$,  
 depend on a single parameter $a$,
to which we will refer as the coupling constant. We will only discuss the regime of relatively small $a$.
Note further, that its
$O(a)$ term have different powers of the coordinate, $(2a)(x^3-x)$, so  $a$-dependence
of the action cannot be absorbed by standard redefinition of the coordinate, as is the case in several examples  discussed previously. 

 In section \ref{sec_PS} we will start with arbitrary coefficients
 of these terms. Calculating
 the first perturbative correction
 by standard perturbative series,
 we will show that for this Hamiltonian the correction $O(a^2)$ cancel each other. Pursuing the issue
  to higher orders, we evaluate
  generic Hamiltonian matrix and show,
  for this case, coefficient cancellations persists to higher orders as well. (We 
   actually checked terms $O(a^4,a^6,a^8)$ and stopped, although it can be continued with larger and larger submatrices.) 

In section \ref{sec_Riccati}
we switch from Schrodinger equation to its Riccati form, and show that
coefficients of the
perturbative series vanish to $all$ orders
(Turbiner, private communication).

The $nonzero$ vacuum energy $E_0(a)$  we first simply calculate
numerically in section \ref{sec_num}, see Fig.\ref{fig_Evac}. Indeed, its observed behavior is very different from that
of generic potentials, for which the lowest
PS terms dominate at small $a$.

Then, following \cite{Behtash:2015zha},
we are looking for paths in the complex plain, classical solutions
of {\em holomorphic Newton’s equations } ( HNE). Unlike those authors, we 
 solve them numerically, thus deriving a  {\em family} of the complex paths with
 all initial directions from their starting point, the global maximum (of inverted potential).
 
 All such paths recoil from some
 complex turning point and come back,
as instanton-antiinstanton pair would do. Following \cite{Behtash:2015zha}, we also call them
{\em complex 
 bions}. All of them happen to possess the same action.  Moreover, the imaginary part of the action is always $\pi$,
so that $exp(-S)$ remains real. 

Finally,
we demonstrate that the vacuum energies obtained from
solutions of Schrodinger equation are in good
correspondence to those actions.

\begin{figure}
    \centering   \includegraphics[width=0.95\linewidth]{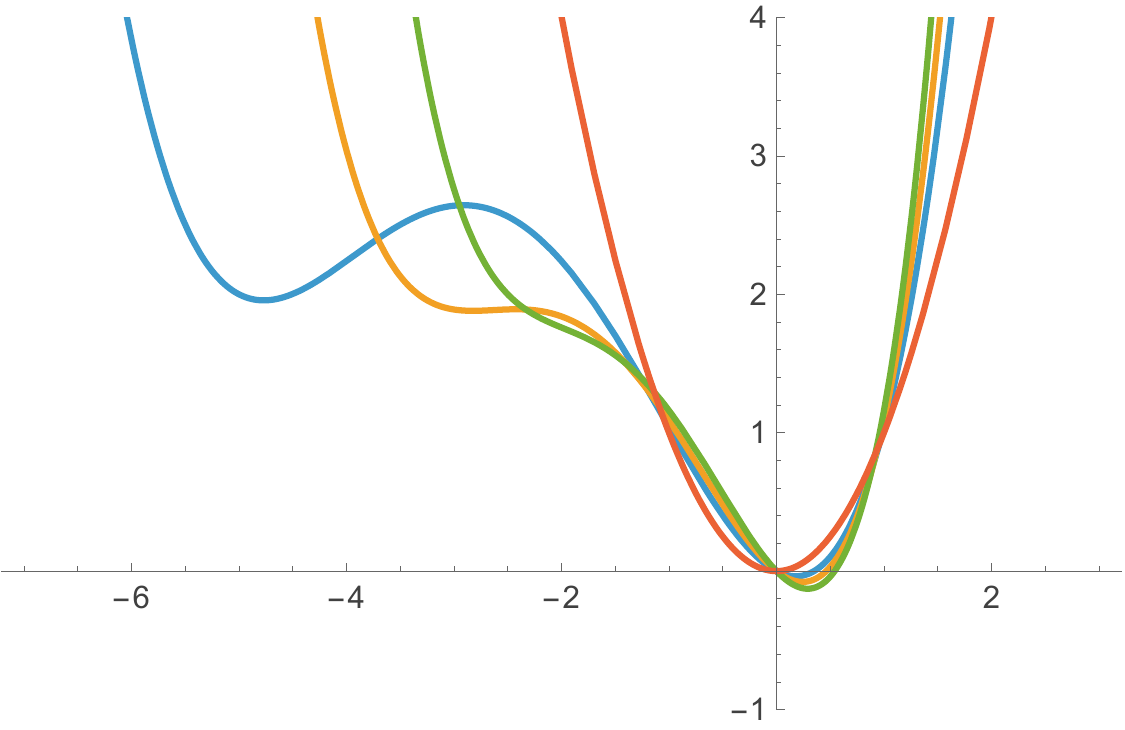}
    \caption{Potentials for $a=0,0.2,0.3,0.4$ (red,blue, orange,green).}
    \label{fig_V}
\end{figure}

\section{Perturbative series} \label{sec_PS}
\subsection{Correction to vacuum energy of order $a^2$}
At $a=0$ the problem (\ref{eqn_pot})
it is just a harmonic oscillator, 
with spectrum given by even integers $E_n=0,2,4...$.

Let us start  using parameter $a$  as a
perturbative coupling.
Explicit calculation of coefficient of  standard perturbation theory is straightforward. The $a^2x^4$ term
appears already in the first order,
and $2a(x^3-x)$ in the second order.
The results are
\ba 
a2_{x^4}&=&\langle 0 |a^2 x^4 |0 \rangle  = (3/4) a^2\\
a2_1 &=&{|\langle 0 |2a(x^3-x) |1 \rangle|^2 \over E_0-E_1}  =  -(1/4) a^2 \nonumber \\
a2_3&=&{|\langle 0 |2a(x^3-x) |3 \rangle|^2 \over E_0-E_3} =-(1/2)a^2  \nonumber \\
\rightarrow a2 &=&\sum_i a2_{i}=0
\ea
Adding them together, one finds that the second order perturbative coefficient $O(a_2)$ is indeed zero.

\subsection{Using the Riccati form of equation, to all orders} \label{sec_Riccati}
 
Looking for higher order coefficients
it is convenient to
follow Turbiner and
 switch to Riccati form of Schrodinger equation 
\ba \psi \sim exp(-\phi(x)),\,\,\,
y(x)=\phi'(x)) \nonumber \\
y(x)^2-y'(x)=V-E 
\ea
Let us now look for a solution for the ground state energy and $y(x)$ in form of perturbative series, 
\be E_0=\sum e_n a^n,\,\,\, y(x)=\sum y_n(x) a^n \ee
Putting those  into Riccati  equation above and separating powers of $a$, we got
set of  equations 
\be E_0=0,\,\,\,\, y_0=x\ee
\be y'_1-2xy_1=E_1-2x^3+2x\ee
\be y'_2-2 x y_2=E_2-x^4-y_1^2 \ee
and for $n>2$
\be y'_n-2xy_n=E_n+\sum_1^{n-1}y_i y_{n-i} 
\ee
Since in the zeroth order the potential is an oscillator, symmetric
in $x\rightarrow -x$, it is clear that $O(a)$ terms in the potential
are both odd and thus their average value must be zero, so $E_1=0$.
The  equation gives  then $y_1=x^2$.

The second eqn has in general a solution 
$$ y_2=e^{x^2}\big[c_2+E_2/2\sqrt{\pi} Erf(x) \big]$$
However, as we have already shown above from ordinary perturbation theory, 
for the potential in question $E_2=0$
due to cancellation of contributions, and the first term
would not give normalizable wave function
. It then forces vanishing of the hole function  as well, $y_2(x)=0$.

The r.h.s. of the third equation contains in general a new term $y_1 y_2$ which in our case  vanishes since $y_2(x)=0$. Then the eqn for $y_3$ is the same as
for $y_2$, and therefore it leads to the same conclusion, $E_3=0,y_3=0$.
This situation continues then to all orders, so one concludes that
perturbation theory for all $E_n,y_n$ terminates. 

If so, then $y=x+a x^2$ and one comes to
a solution of Schredinger equation
\be \psi\sim exp(-x^2/2-x^3/3) 
\ee
which is simple but, unfortunately,
$unphysical$ since it is $non-normalizable$. (In fact, starting from it Turbiner invented the discussed  potential (\ref{eqn_pot})). 

\section{The actual vacuum energy $E_0(a)$}
\label{sec_num}
Completing discussion of perturbative series, we return
to study of the problem in full.
The shape of the potential (\ref{eqn_pot}) is shown in Fig.\ref{fig_V}. One can see that
there are two regimes: at small enough $a$ there is a second minimum, which disappear above
some critical value.

Since derivative of our potential is a polynomial of power 3, in the complex plane there are always three extrema. Analytic expressions for
their location can be easily obtained e.g. by Wolfram Mathematica,
but we skip those as they are lengthy and not very relevant.

\begin{figure}[t]
    \centering
    \includegraphics[width=0.95\linewidth]{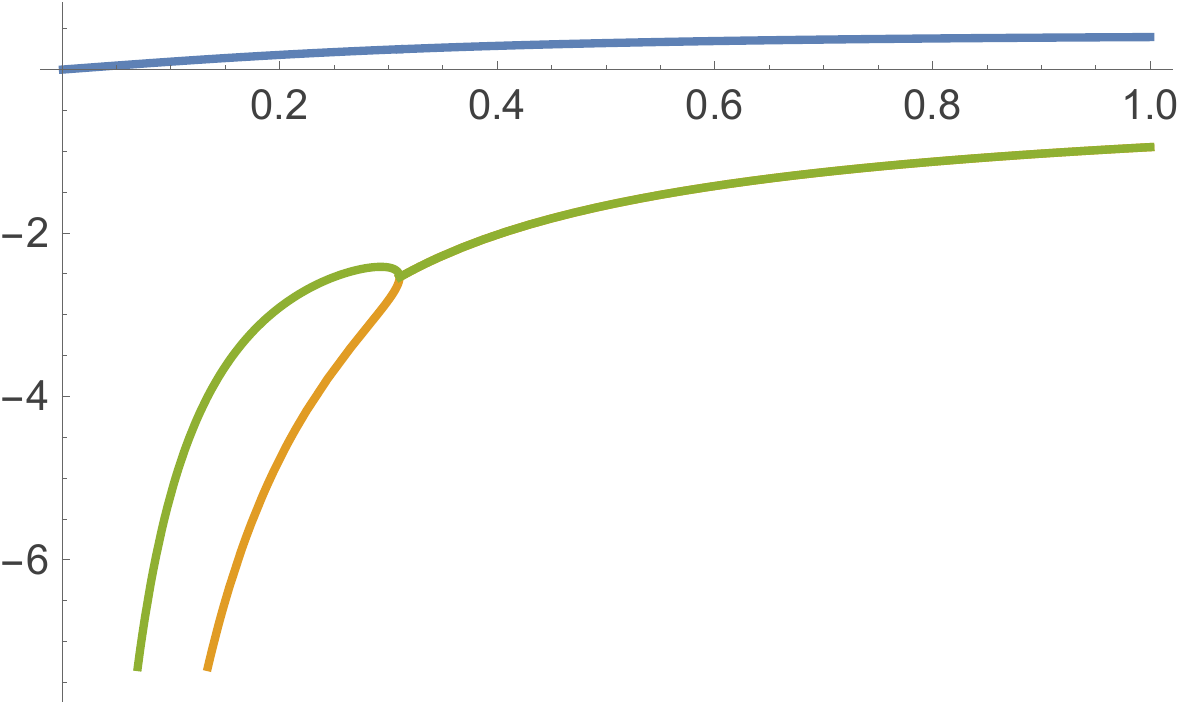}
        \includegraphics[width=0.95\linewidth]{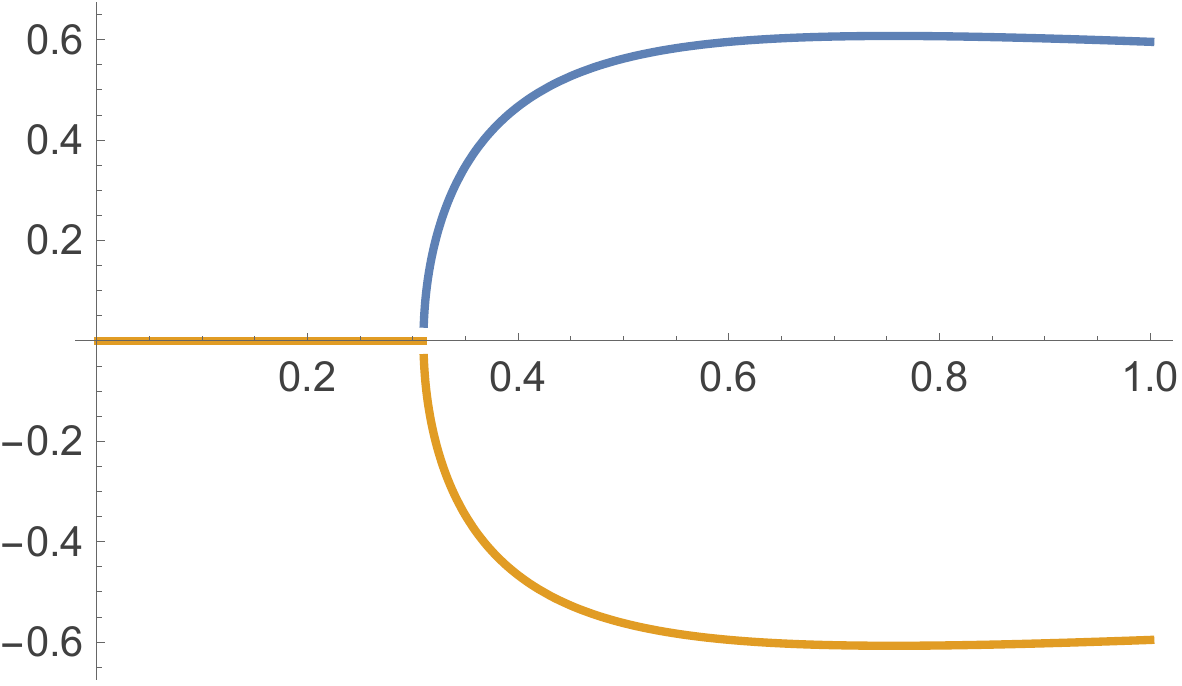}
    \caption{Real and imaginary parts of the extrema, as a function of parameter $a$.
    Lower plot to complex conjugate pair only.}
    \label{fig_extrema}
\end{figure}

Those  one can clearly see
in Fig.\ref{fig_extrema}. For small $a<a_c$ all
three extrema are real, then they collide, and above $a_c$   two are complex conjugates (see the lower plot)

\begin{figure}[h!]
    \centering       \includegraphics[width=0.95\linewidth]{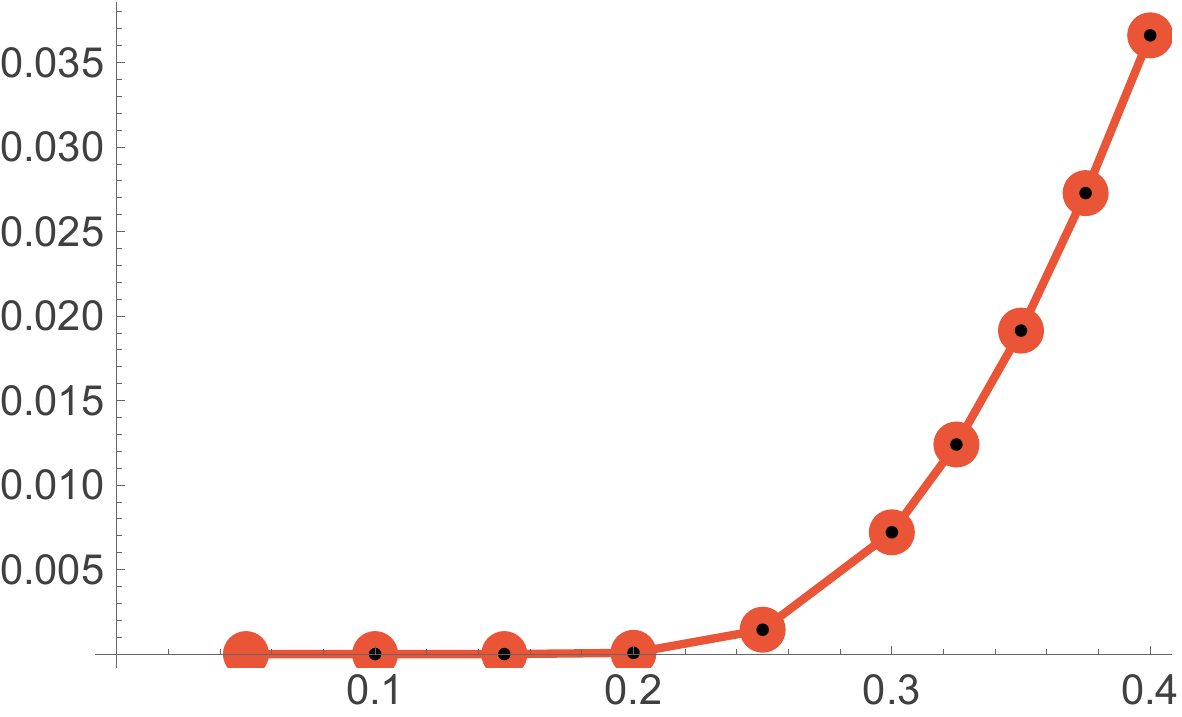}     \includegraphics[width=0.95\linewidth]{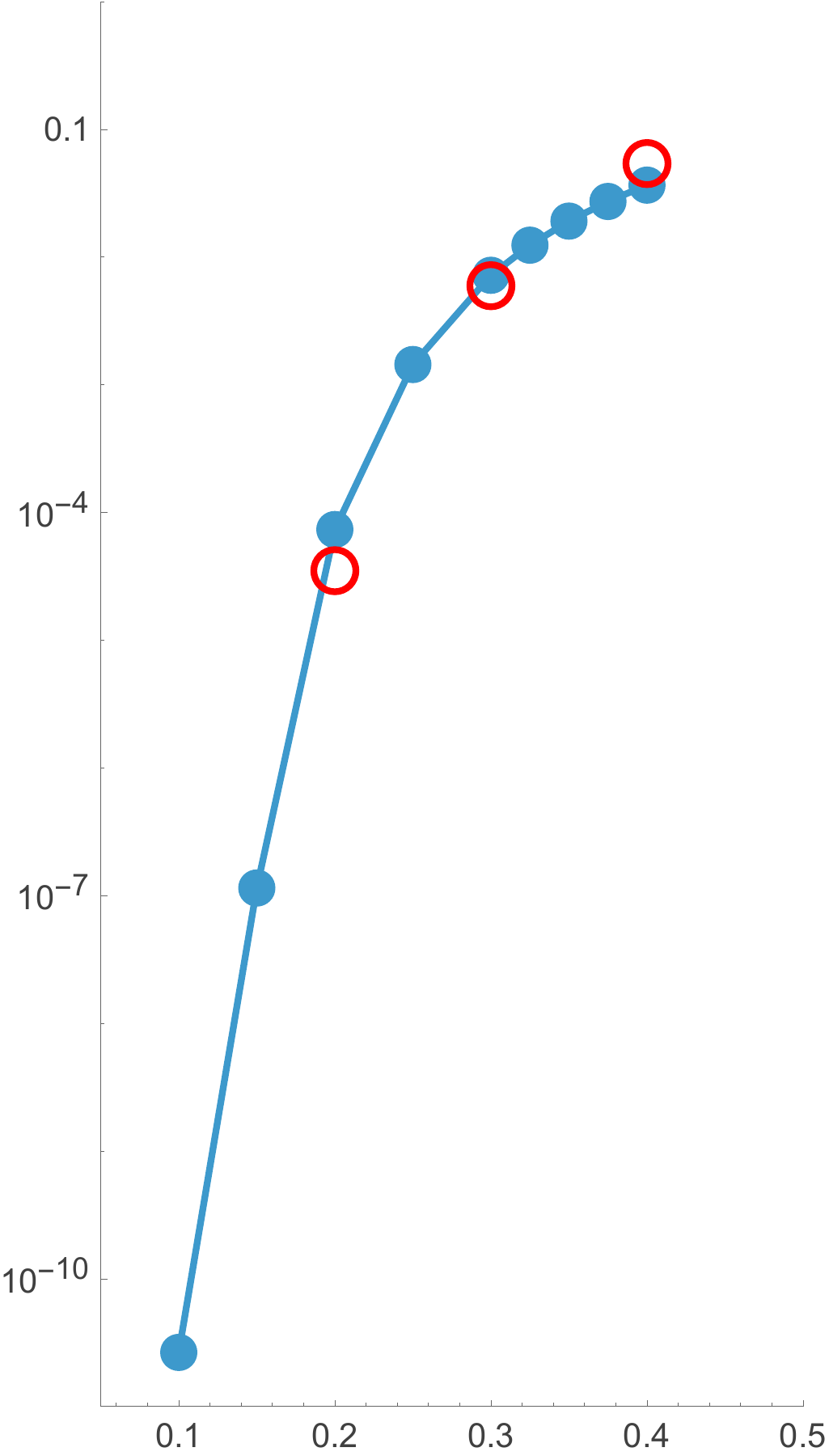} 
    \caption{The ground state (vacuum) energy versus parameter $a$. The upper (red) and lower  (blue) points
    are the same data, just plotted in linear and logarithmic plots. Red circles
    on the lower plot represent effects of complex bions according to (\ref{eqn_contribution_bions}) derived below.}
    \label{fig_Evac}
\end{figure}

There is no problem to find the
spectrum at any values of $a$ numerically. It is indeed nonzero,
and its
dependence on $a$
is shown in
 Fig.\ref{fig_Evac}. As shown in the upper plot, deviations of vacuum energy from zero
 are invisible at small $a$. They are  seen in
 the lower (logarithmic) plot, which has to be obtained
 with higher accuracy. 
 
 Such dependence on
 $a$ does not occur
  for generic potentials which have perturbative corrections, e.g. 
 of the lowest order $O(a^2)$.
This plot clearly indicate that
vacuum of Turbiner potential is
completely nonperturbative, 
described by some version
of ``instanton ensemble".

%
%

\section{A family of complex bions }
There is a well known ``bounce"
real path starting from the lower
maximum (of inverted potential -V).
But, since  it does not touch 
the main global maximum, we also do not
care about two different regimes in $a$, with different arrangements
of two other extrema.

Relevant solutions to HNE 
$$ m{d^2 z \over dt^2}={\partial V(z(t)\over dz}$$
should all start from the global maximum of $-V$ and move into the complex plane, where it may find a turning
point and return back, closing the solution. 

\begin{figure}[h!]
    \centering
    \includegraphics[width=0.95\linewidth]{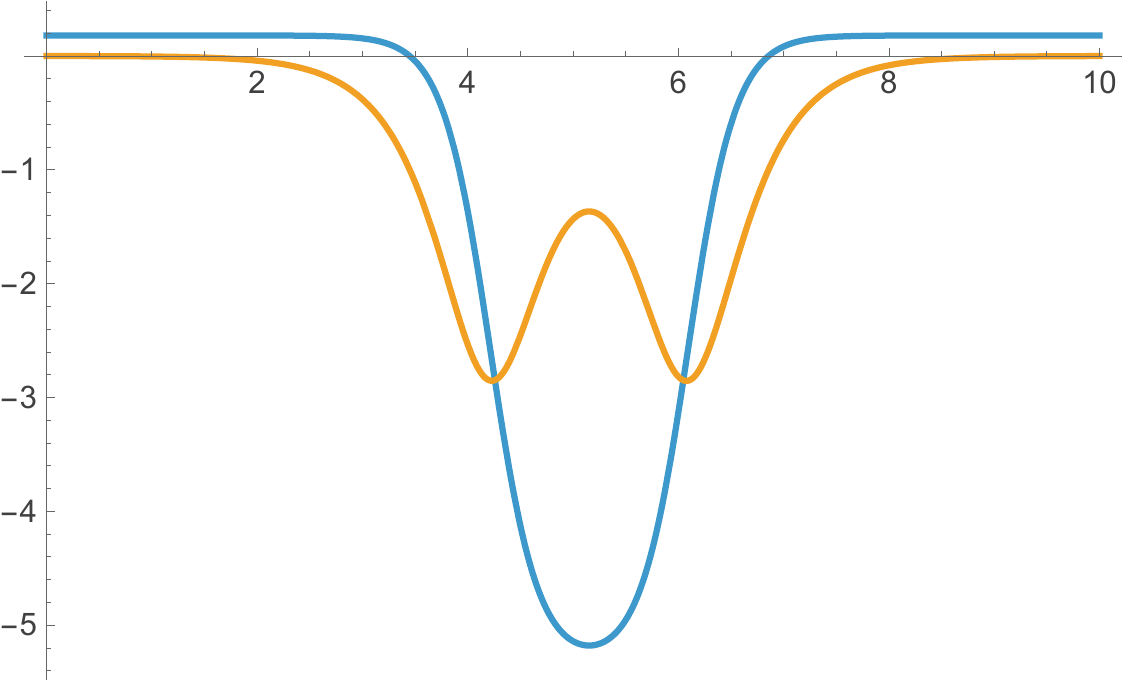}
    \caption{Compex bion, a=0.2, shows $X(t)$ and $Y(t)$ as a function of ``time" $t$.}
    \label{fig_bion}
\end{figure}

Since we  proceed numerically, we may 
 start from the global maximum in any direction. This is done by small displacement of the initial point  by $\delta (cos(\phi),sin(\phi)$,  typically $\delta\sim 10^{-3}$. The angle  $\phi$ is an arbitrary angle on which  complex bion solution depends. 
At $a=0.2$ an example of it with $\phi=-\pi/2$
(starting along the imaginary axes) is shown
in Fig.\ref{fig_bion}

\begin{figure}[h!]
    \centering        \includegraphics[width=0.95\linewidth]{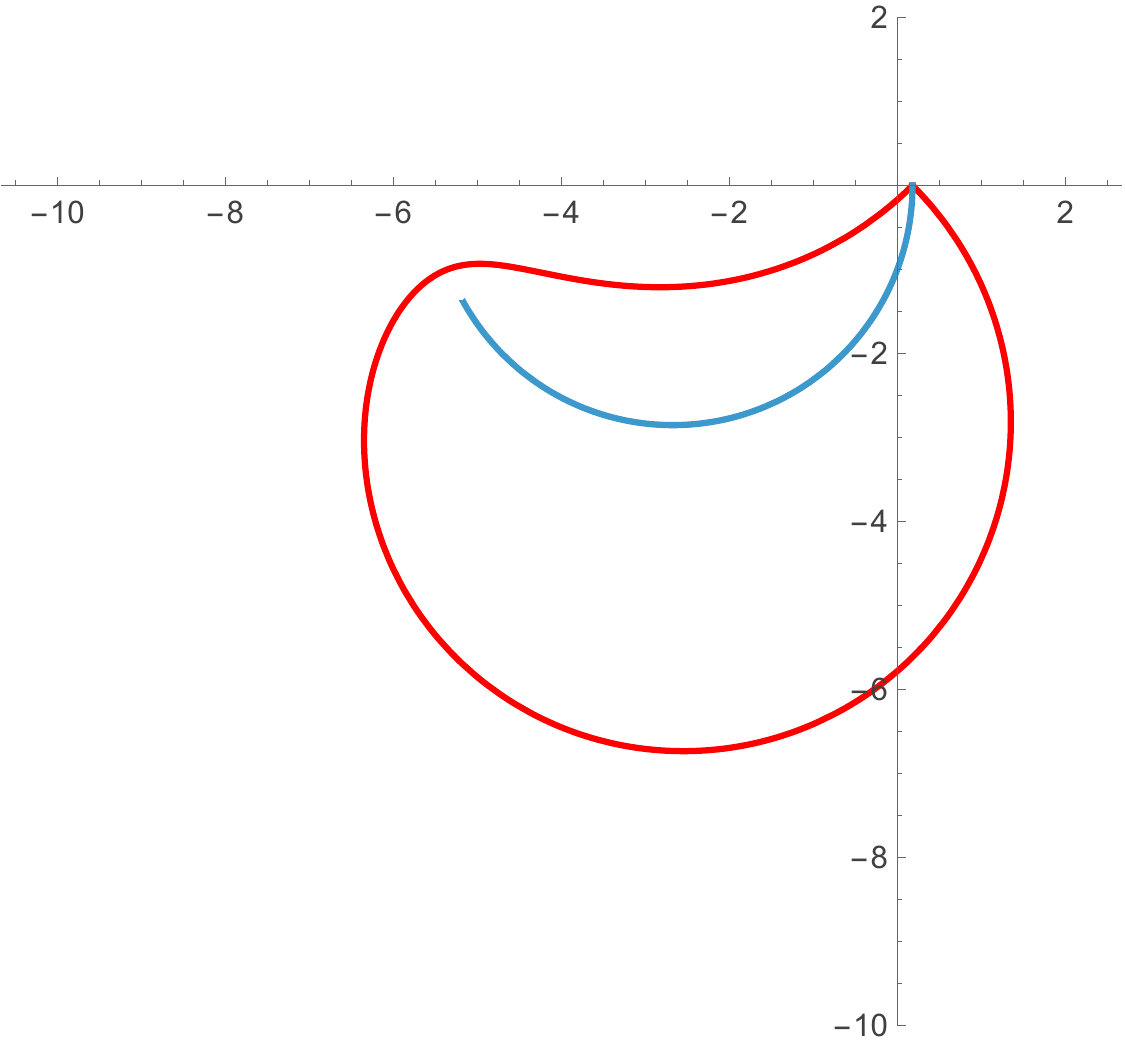}
    \caption{Compex bions for $a=0.2$. The parametric plot shows complex $X,Y$ plane with parameter $t$. Blue curve is 
    passed twice, it is for angle $\phi=-\pi/2$. The red curve is for $\phi=-3\pi/4$ (or  $\phi=-\pi/4$, same curve in opposite direction). }
    \label{fig_bions}
\end{figure}

As a result, we find that holomorphic Newton’s equation has a family of complex bion solutions, parameterized by their
initial angle $\phi$. Not surprisingly,
all of them have the same action. For our
examples of bions 
\ba S_{kin}(a=0.2)=\int dt{1\over 4}( z'(t))^2 =   6.68618 + 1.57044 i, \nonumber \\
S_{pot}(a=0.2) =\int dt V(z(t))=6.23949 + 1.57045 i \ea
 The shape of these solutions is better shown in parametric plot Fig.\ref{fig_bions}. The numerically calculated total action is thus
\be S_{tot}(a=0.2)=12.9246 +i\cdot 3.14151  \ee
It is the same for any shape of the path
(within numerical accuracy) for any value of the 
initial angle $\phi$. (Due to holomorphic nature of the equation
 it is of course
is just another manifestation of Cauchy theorem, allowing deformations of the contour.)

Note that while the action has nonzero imaginary part,  the amplitude
$exp(-S)$ remains real, 
since the imaginary part happen to be exactly $\pi$, and $exp(\pm i\pi)=-1$.
Bions going into upper half plane
are complex conjugate.

Originally found as a numerical curiosity, this observation  has already been explained in 
\cite{Behtash:2015zha}. This result can be traced to a limit in which 
a Cauchy contour is increased  into a circle of large radius, 
where it becomes defined by the asymptotics
of the potential,  given in terms of the highest powers 
of the coordinate $z$ in the potential. 

\begin{figure}[h!]
    \includegraphics[width=0.85\linewidth]{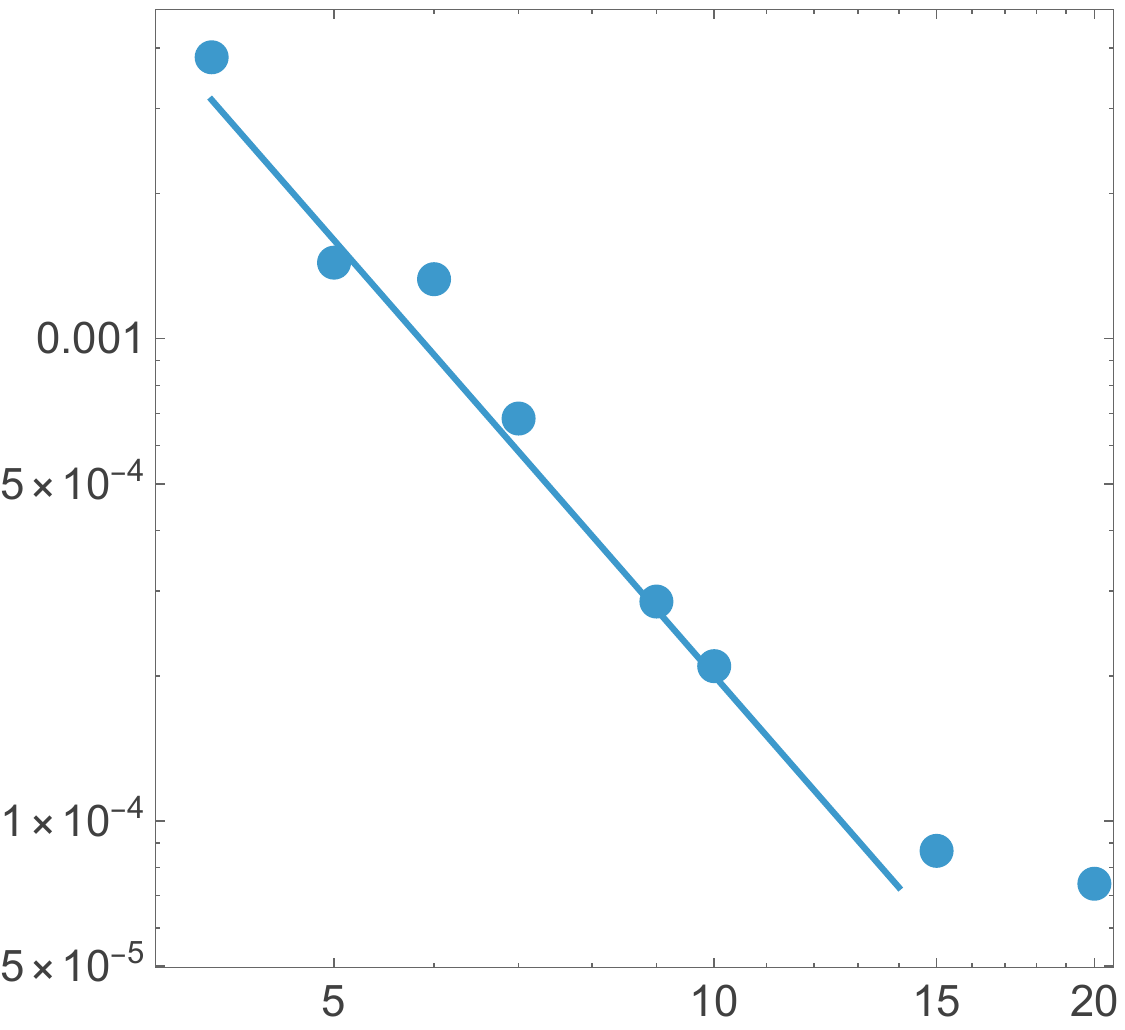}
        \includegraphics[width=0.85\linewidth]{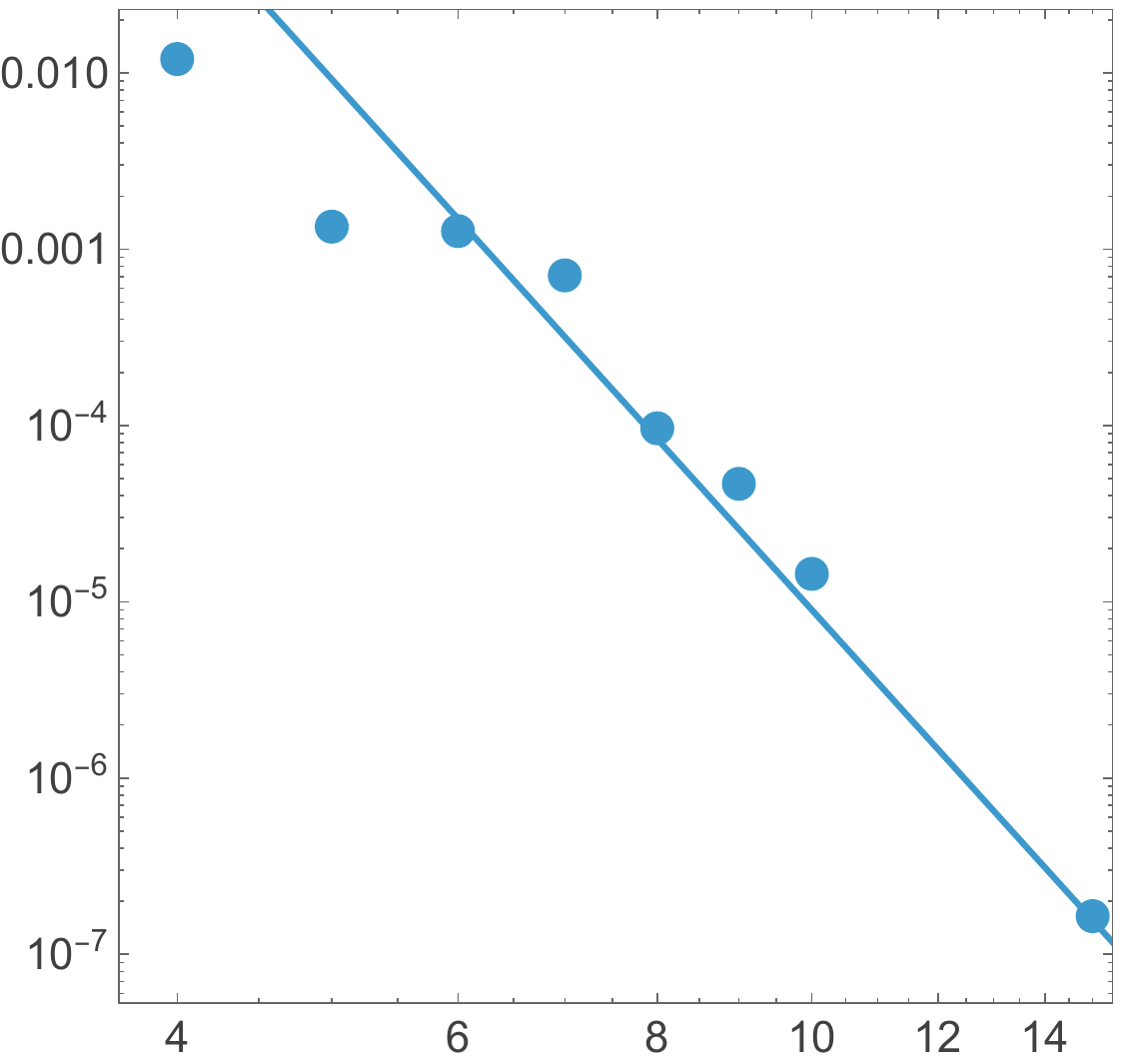}
    \caption{Points are the ground state energy for $n\times n$ matrix  versus $n$. The upper plot is for $E_0(n,a)$ for $a=0.2$.  The lower plot
    shows $deviation$ $E_0(n,a)-E_0(\infty,a)$
    for $a=0.4$. Line in the upper plot is $\sim n^{-3}$, and in the lower $\sim n^{-10}$, shown  for comparison.}
    \label{fig_E02RG}
\end{figure}

Finally, it is tempting to relay the obtained
bion actions with the vacuum energy 
calculated earlier and shown in Fig.\ref{fig_Evac}.

Of course, we do not yet 
know the determinant of the fluctuation operator, only the classical actions. Let us 
evaluate the vacuum energy as
\be \label{eqn_contribution_bions}
E_{bions}(a) =C_{det} \sqrt{Re(S)} exp(-Re(S))\ee
The square root is expected to come from the Jacobian of a single
 zero mode. The constant $C_{det}$ represents an unknown fluctuation determinant. Three red circle in Fig.\ref{fig_Evac} are calculated from it, for crudely fitted value $C_{det}=4$. The 
qualitative  behavior is clearly captured.
Of course, to get really accurate description one needs
to calculate the two-loop $O(1/S)$, and  perhaps even
three loop corrections $O(1/S^2)$, as we 
did for DW potential in \cite{Escobar-Ruiz:2015nsa}.
\section{Hamiltonian Renormalization Group}
Renormalization group (RG) equation by Gell-Mann and Law  is based on a relation between the
 cutoff scale and the value of the running coupling. Its
 discretized form was famously used for second order phase transitions by K.Wilson  using $\epsilon$ expansion near space dimension $d=4$.

Yet this paper uses a different form of it known as
Hamiltonian RG (HRG) by Glazek and Wilson \cite{Glazek:1993rc}, based on systematic widening 
of the size of Fock space of states, in which eigensystem of some Hamiltonian is defined.
Representation of Hamiltonians as finite-size
matrices in some basis is a very popular 
way of solving spectroscopic quantum problems
in atomic, nuclear, particle physics, in general in
multiple quantum manybody problems. 
Yet
systematic studies of accuracy and convergence 
of such approaches are in most cases
 missing.

\begin{figure}
    \centering
    \includegraphics[width=0.85\linewidth]{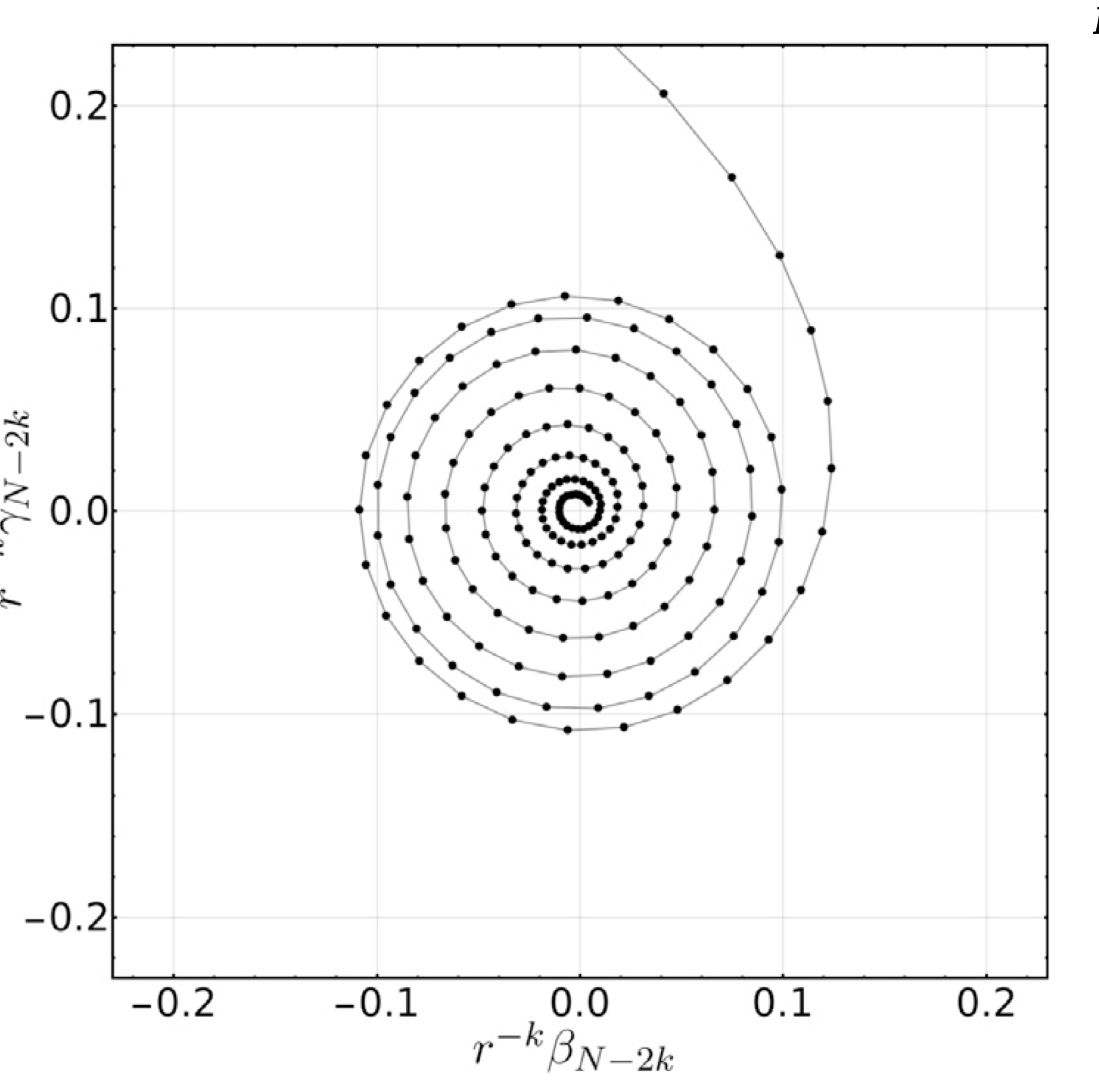}
    \caption{RG evolution near the attractive nearly-fixed point  for the running energy cutoff $N-2k$ and $k$ ranging from 8 to 200.}
    \label{fig_spiral}
\end{figure}

Among  basic studies is  in particular that of quantum anharmonic oscillator 
\cite{spiralquartic}. It was found  there  that the HRG approach to the fixed point happens via some $spirals$, see e.g. Fig.\ref{fig_spiral} from it.

\begin{widetext}
It is a straightforward exercise to calculate
Hamiltonian matrix of $H_T$ elements in harmonic basis and diagonalize it.
For example, here are $10\times10$ part of it
\begin{scriptsize}
\begin{tabular}{|cccccccccc|}        
${3 a^2\over 4}$ & ${a\over \sqrt{2}}$ & 
${3 a^2 \over \sqrt{2}}$ & $\sqrt{3} a$ &  $\sqrt{3\over2} a^2$ & 
0&  0& 0& 0& 0 \\
 $ {a\over\sqrt{2}}$ & 2 + ${15 a^2 \over 4}$ & 4 a& 5 $\sqrt{3\over 2} a^2$ & 
  $2 \sqrt{3}$ a & $\sqrt{15\over 2} a^2$ &
  0 & 0 & 0 & 0 \\
  ${3 a^2\over\sqrt{2}}$ & 4 a& 
  $4 + {39 a^2\over 4}$ & 7 $\sqrt{3 \over 2}$ a& 7 $\sqrt{3} a^2$ & $\sqrt{30}$ a& 
  3 $\sqrt{5 \over 2} a^2$ & 0& 0& 0 \\
  $\sqrt{3}$ a& 5 $\sqrt{3\over 2} a^2$ & 
  7 $\sqrt{3 \over 2}$ a & 6 + $75 a^2/4 $ & $10 \sqrt{2}$ a & 9 $\sqrt{5} a^2$ & 
  2 $\sqrt{15} a$ & $\sqrt{105 \over 2} a^2$ & 0 & 0 \\
  $\sqrt{3 \over 2} a^2$ & 2 $\sqrt{3}$ a& 
  7 $\sqrt{3} a^2$& 10 $\sqrt{2}$ a& 8 + $123 a^2/4 $ & 13 $\sqrt{5 \over 2}$ a& 
  11 $\sqrt{15 \over 2} a^2$ & $\sqrt{105}$ a& $\sqrt{105} a^2$ & 0\\
  0& 
  $\sqrt{15\over 2} a^2$ & $\sqrt{30}$ a& 9 $\sqrt{5} a^2$ & 13 $\sqrt{5 \over 2}$ a & 
  10 + $183 a^2/4 $& 16 $\sqrt{3}$ a& 13 $\sqrt{21 \over 2} a^2$ & 2 $\sqrt{42}$ a& 
  3 $\sqrt{21} a^2$\\
  0& 0& 3 $\sqrt{5 \over 2} a^2$ & 2 $\sqrt{15}$ a& 
  11 $\sqrt{15 \over 2} a^2$ & 16 $\sqrt{3}$ a& 12 + $255 a^2/4 $ & 19 $\sqrt{7 \over 2}$ a & 
  15 $\sqrt{14} a^2$ & 6 $\sqrt{7}$ a\\
  0& 0& 0& $\sqrt{105 \over 2} a^2$ & 
  $\sqrt{105}$ a& 13 $\sqrt{21 \over 2} a^2$ & 19 $\sqrt{7 \over 2}$ a& 14 + $339 a^2/4 $ & 
  44 a& 51 $\sqrt{2} a^2$\\ 0& 0& 0& 0& $\sqrt{105} a^2$ & 2 $\sqrt{42}$ a& 
  15 $\sqrt{14} a^2$ & 44 a& 16 + $435 a^2/4$ & 75 a/$\sqrt{2}$ \\
  0& 0& 0& 
  0& 0& 3 $\sqrt{21} a^2$ & 6 $\sqrt{7} a$ & 51 $\sqrt{2} a^2$ & (75 a)/$\sqrt{2}$ & 
  18 + $543 a^2/4 $\\
    \end{tabular}
    \end{scriptsize}
\end{widetext}
Its determinant is a polynomial proportional to certain powers
of $a$, because at $a\rightarrow 0$ the ground state
energy is zero. This power grows with the matrix size.
For example for matrix of $4\times 4$ it is
$$ det(H_{44})=120 a^4 + (1755 a^6)/8 + (23625 a^8)/256 $$
starting from $a^4$ and not $a^2$ as one could anticipate. For the  $10\times 10$ sub-matrix
given  above the determinant starts with power $a^8$. What it means is that corrections $\sim a^2,a^4,a^6$ to the ground state energy has already cancels
inside this matrix, while terms $\sim a^8,...$ are present because cancellations not yet happened inside such sub-matrix.

In  Fig.\ref{fig_E02RG} we show how the exact ground state energies are approached 
when the size $n$ of the set of included states (Hamiltonian in $n\times n$ form)  is
systematically increased. There is no spiral, with the RG evolution 
 converging to exact values monotonously and quite rapidly.

{\bf Summary}: we studied an example of a potential (\ref{eqn_pot})
suggested by Turbiner. By different means we show that its perturbative series are vanishing to all orders. We show it directly, without usage of any fermions and
supersymmetry.

The nonvanishing vacuum energy is calculated numerically for all couplings $a$. Its dependence on $a$ is clearly nonperturbative. We give evidences that it is due to
``instanton-antiinstanton correlated pairs",
in form of the ``complex bion" classical solutions to HNE, found numerically. 

We also studied how the results are approached when
$n\times n$ part of the Hamiltonian is used.
What is found, the approach in this case is very rapid and show no sign of oscillations.

As an outlook, it would be interesting
to see how our results correspond
to different versions of ``resurgence" relations, and in
particular whether using them one can
find the series in higher order fluctuations around such complex bions.

It is also tempting to ask if similar complex classical paths can
be found for QFT problems, such as for the gauge 4d theories.

{\bf Acknowledgements:} This paper
has originated from  conversations and specific suggestions by Alexander Turbiner,
which are all greatly appreciated.
My work in general is supported by
DOE Office of science, under Contract  No. DE-FG-88ER40388.

\bibliography{main.bib}
\end{document}